# Stripe charge order driven manipulation of Majorana bound states in 2M-WS$_2$ topological superconductor

Xuemin Fan[1,2,3†], Xiaoqi Sun[4†], Penghao Zhu[4†], Yuqiang Fang[5,6†], Yongkang Ju[7], Yonghao Yuan[1,2,3], Fuqiang Huang[5,6*], Taylor L. Hughes[4], Peizhe Tang[7,8*], Qi-Kun Xue[1,2,3,9,10*] and Wei Li[1,2,3*]

[1]*State Key Laboratory of Low-Dimensional Quantum Physics, Department of Physics, Tsinghua University, Beijing 100084, China*

[2]*Collaborative Innovation Center of Quantum Matter, Beijing 100084, China*

[3]*Frontier Science Center for Quantum Information, Beijing 100084, China*

[4]*Institute for Condensed Matter Physics and Department of Physics, University of Illinois at Urbana-Champaign, Urbana, Illinois 61801, USA*

[5]*State Key Laboratory of High Performance Ceramics and Superfine Microstructure, Shanghai Institute of Ceramics, Chinese Academy of Science, Shanghai 200050, China*

[6]*State Key Laboratory of Rare Earth Materials Chemistry and Applications, College of Chemistry and Molecular Engineering, Peking University, Beijing 100871, China*

[7]*School of Materials Science and Engineering, Beihang University, Beijing 100191 China*

[8]*Max Planck Institute for the Structure and Dynamics of Matter, Center for Free-Electron Laser Science, 22761 Hamburg, Germany*

[9]*Beijing Academy of Quantum Information Sciences, Beijing 100193, China*

[10]*Southern University of Science and Technology, Shenzhen 518055, China*

[†]These authors contributed equally to this work.
[*]To whom correspondence should be addressed: huangfq@mail.sic.ac.cn; peizhet@buaa.edu.cn; qkxue@mail.tsinghua.edu.cn; weili83@tsinghua.edu.cn

Majorana bound states (MBSs) are building blocks for topological quantum computing[1,2]. They can be generated via the combination of electronic topology and superconductivity[3-20]. To achieve logic operations via Majorana braiding, positional control of the MBS must be established. To this end, exotic co-existing phases or collective modes in an intrinsic topological superconductor can provide a tuning knob to manipulate the MBS. Here we report the observation of a striped surface charge order coexisting with superconductivity and its controllable tuning of the MBS in the topological superconductor $2M$-$WS_2$ using low-temperature scanning tunneling microscopy. By applying an out-of-plane magnetic field, we observe that MBS is absent in vortices in the region with strong stripe order. This is in contrast to adjacent underlaying layers without charge order where vortex-bound MBSs are observed. Via theoretical simulations, we show that the surface stripe order does not destroy the bulk topology, but it can effectively modify the spatial distribution of MBS, i.e., it pushes them downward away from the $2M$-$WS_2$ surface. Our findings demonstrate that the interplay of charge order and topological superconductivity can be used to manipulate the positions of the MBS, and to explore of new states of matter.

Topological materials can host exotic Majorana bound states (MBSs) that obey non-Abelian statistics required for topological quantum computing. Evidence of MBS has been observed in topological insulator-superconductor hybrid structures[5-8,17,18,20]. Possible MBSs have also been reported in the vortex cores of intrinsic Fu-Kane topological superconductors[3], such as $FeSe_{0.5}Te_{0.5}$ (ref. [13,21]), $Li_{0.84}Fe_{0.16}OHFeSe$ (ref. [14]), $2M$-$WS_2$ (ref. [15]) and LiFeAs (ref. [22]), in which bulk superconductivity coexists with topological surface states. Recently, there has been emerging interest in exotic electronic order in intrinsic topological superconductors. Such systems offer a platform to manipulate the MBS and study the interplay between the MBSs and the exotic co-existing order. Indeed, a bulk charge density wave (CDW) in strained LiFeAs can renormalize the bulk electronic band structure without changing its topological properties, and can pin vortices to generate arrays of MBS[23]. Alternatively, since the MBSs in Fu-Kane superconductors live on the surface, modifications of the surface can be used to control the position and distribution of MBSs. Furthermore, it is of fundamental interest to study the interplay between surface electronic order and MBSs. As a first step towards understanding the complex interactions between the MBS and other co-existing electronic order, we report the observation of surface charge order in an intrinsic topological superconductor $2M$-$WS_2$ and find that it can be used to manipulate the MBS.

**Stripe charge order in $2M$-$WS_2$**

We use $2M$-$WS_2$ single crystals, in which topological surface states, superconductivity and MBS have been observed[15,24-27]. The material $2M$-$WS_2$ is a van der Waals crystal constructed from $1T'$-$WS_2$ monolayers along the $c$-direction (Fig. 1a). Within each monolayer, the deviation of W atoms from the $[WS_6]^{8-}$ octahedral centre leads to a zigzag structure along the $a$-direction (see Fig. 1a). The atomically resolved scanning tunneling microscopy (STM) image of $WS_2$ (Fig. 1b) shows this zigzag structure (marked by dashed green lines), consistent with the structure in Fig. 1a. The large diagonal features running from bottom left to top right in Fig. 1c correspond to moiré patterns, forming in the cleaving process of the single crystal, during which the topmost layer of WS$_2$ is distorted with a small twist angle. The moiré patterns will be discussed in detail in Discussion section.

In addition to the zigzag atomic chains, we also observe long-range electronic stripe modulations in STM topographic images (denoted by yellow arrows in Fig. 1b and c). These stripes break the rotational and translational symmetries of the lattice and have an incommensurate spatial period of 1.31 nm with slight fluctuations (1.24-1.38 nm) in different regions of the sample. Moreover, in contrast to that of ordinary charge order previously revealed by STM[28,29], the orientation of the stripes observed here is not uniform, and the local distortions in the orientation generate a line segment-like feature (rather than spot-like feature) centered at the wave vector $q_0$ in the fast Fourier transform (FFT) image as shown in Fig. 1d. We note that $q_s$ and $q$ in Fig. 1d correspond to the wave vectors of the S-S lattice and the zigzag chain, respectively, and they still manifest as spot-like features. To highlight the stripe modulations in Fig. 1c, we perform inverse fast Fourier transform (IFFT) to $q_0$ and the distorted stripe patterns are clearly shown in Fig. 1e. Scanning tunneling spectroscopy (d$I$/d$V$ spectrum) measures the local electronic density of states (DOS), while d$I$/d$V$ mapping probes the spatial distribution of DOS at specific energies over a certain area[30,31]. Figure 1f shows the energy dependent line-cut of the stripes extracted from a series of d$I$/d$V$ mappings at the same location. The period of the stripes is unchanged at different energies, indicating that the observed stripes are static electronic modulations rather than quasiparticle interference patterns. Such charge modulation extends to the energy range of -20 meV to 50 meV.

We then investigate the relationship between the superconductivity and the stripes. Low-energy-scale d$I$/d$V$ spectra taken at 400 mK exhibit U-shaped superconducting gaps, indicating the nodeless superconducting pairing in 2M-WS$_2$ (Fig. 1g). Such an observation is consistent with previous work[15]. In addition, the superconducting coherence peaks show spatially periodic modulations. The periodic modulations of the coherence peaks are closely related to the corresponding locations where the spectra are acquired (inset, Fig. 1g). The spectra with lowest superconducting coherence peaks are always taken at the valleys of the stripe patterns (yellow arrows in Fig. 1g). The periodic evolution of averaged heights of the coherence peaks at different locations is summarized in Fig. 1h, demonstrating the intimate correlation of superconductivity and stripe charge order.

**The absence of MBS in the stripe region**

We will now show that the distribution of magnetic vortices exhibits intriguing dependence on the stripes. Two types of stripes with different intensity, namely strong stripes and weak stripes, are observed (Figs. 2a and b). The tunneling spectrum in the strong stripe region shows a gap-like feature with suppressed density of states near $E_F$ compared with that in the weak-stripe region (Fig. 2c). The stripes across the boundaries of the two stripe regions (the dashed lines in Fig. 2) are continuous. We utilize two criteria to distinguish the two regions. (1) The modulation strengths of the stripes in the two regions are different. Figure 2d and e present the fast Fourier transform results of the strong and weak stripe regions, respectively. The intensities of the two FFT images are normalized by the zigzag signal $q$. The intensity of $q_0$ in the strong stripe region is significantly higher than that in the weak stripe region, directly evident in the line-cut profiles crossing $q_0$ (Fig. 2f). (2) The d$I$/d$V$ spectra of the two regions are distinct from each other (Fig. 2c). The density of states near $E_F$ are quite different in the two regions, resulting in a distinct contrast in the d$I$/d$V$ mapping taken at 5 meV (Fig. 2b). The boundaries between the strong and weak stripe regions can therefore been easily determined in Fig. 2b.

Interestingly, the distribution of magnetic vortices in the two regions exhibits a large distinction. Figure 2g-j shows the magnetic-field dependence of the vortices. The vortices exist in only the strong stripe regions, and their density increases with increasing magnetic fields. The vortices in the strong stripe regions start to merge into each other at ~1.2 T (see Fig. 2i). In contrast, no magnetic vortices pass through the weak stripe regions. The absence of vortex here might be due to the small domain size of the weak stripe regions[32]. The superconductivity in the weak stripe regions transits to the normal state at 1.8 T (Fig. 2j). The estimated $H_{C2}$ of the strong stripe region is 1.6 T (see Extended Data Fig. 1). The $H_{c2}$ is inhomogeneous over the sample due to the existence of local strain, giving rise to the unregular distribution of vortex at 1.2 T (Fig. 2i). We note that magnetic fields up to 12 T do not change the distribution of the stripes (see Extended Data Fig. 2), indicating a non-magnetic origin of the stripes.

Bulk 2M-WS$_2$ hosts a topologically non-trivial electronic structure[24,33] with a non-trivial $Z_2$-invariant, as observed in ARPES measurements[25]. Below the superconducting transition temperature ($T_c$), 2M-WS$_2$ without surface charge order becomes a Fu-Kane topological superconductor and has an odd number of the MBSs trapped at the centre of vortex cores[15]. Interestingly, in 2M-WS$_2$ with strong surface charge modulation observed here, the MBSs are absent in the vortices. To affirm this, we have checked more than 25 vortices in the region, and a zero-bias conductance map of a typical single vortex is shown in Fig. 2k; note that it is anisotropic and elongated along the $a$-direction. A series of d$I$/d$V$ spectra shown in the false-colour image in Fig. 2l were taken across the vortex along cut 1 of Fig. 2k. The zero-bias conductance peak at the vortex core splits into two symmetric branches in energy as the tip moves away, which can be attributed to Caroli-de Gennes-Matricon bound states (CBSs)[34-36]. Besides the splitting branches, no remaining zero-bias peak coexists with CBSs off the vortex centre in the spectra[15], indicating the absence of MBS in the vortices of 2M-WS$_2$ within the stripe region.

**The appearance of MBS in the stripe-free region**

To understand the effect of the charge order on the MBS, we need to establish its character and universally robust features. To this end, we investigated the charge order in different samples. We confirmed that the stripe modulations develop on only the topmost layer and in certain regions of WS$_2$. Figure 3a shows an area including three WS$_2$ layers, named top layer, middle layer and bottom layer. The step height between two adjacent layers is 6 Å, corresponding to the monolayer thickness of WS$_2$. The top layer is the one that we have shown before, in which the moiré pattern and the stripes are observed. We argue that the moiré pattern may be formed during the cleaving process, in which the topmost layer is distorted with a small twist angle. In contrast, the stripe and moiré patterns are absent in the middle and bottom layers (Fig. 3b and c). A small amount of adatoms are visible in the underlying layers, which might be the residual alkali metal atoms used during the synthesis of 2M-WS$_2$ (ref. [24]). These observations suggest that the striped charge modulation is a surface phenomenon and is localized on the surface. Furthermore, superconductivity is enhanced in the underlying layers with no stripes, where stronger coherence peaks and cleaner superconducting gaps are observed in the d$I$/d$V$ spectra (Fig. 3d).

Remarkably, the suppression of the stripes and the enhancement of superconductivity, allow the MBS to recover in the vortices on the bottom layer. Correspondingly, a zero-bias conductance peak shows up in the d$I$/d$V$ spectrum at the vortex centre (left panel, Fig. 3e). From the d$I$/d$V$ spectra taken across the vortex along the $a$-axis, an apparent three-peak feature appears in the d$I$/d$V$ spectra off the core centre (right panel, Fig. 3e), demonstrating the existence of MBS. In the false-colour image (see the black and red arrows in middle panel and the schematic in the inset of right panel of Fig. 3e), the evolution of the splitting branches and the non-split branch are clearly presented. The appearance of non-split zero-bias peak[15,23] indicates the recovered MBS in the vortex in 2M-WS$_2$ without stripes.

**Discussion**

Now we discuss the possible origin for the stripe order in 2M-WS$_2$ based on our observations. First, the conventional mechanism for charge order induced by the electron-phonon interaction can be excluded. By using density functional theory, we calculate the electronic structures and phonon bands of monolayer, bilayer, and bulk 2M-WS$_2$. We do not observe any negative phonon mode or phonon softening in these materials with the change of carrier doping (Methods and Extended Data Fig. 4 and 5). Second, we speculate that the cleaving process of the single crystal may play an important role for the development of the stripes and the moiré structures (Fig. 4a). The formation of the moiré superstructures on the topmost layer of WS$_2$ may induce local strain and modify the surface electronic structure, giving rise to the development of the stripe order. Stripes and moiré patterns seem to have a symbiotic relationship, although similar stripes appear on the surfaces with different moiré patterns (Figs. 1c and 2a). The moiré patterns look different to each other (Figs. 1c and 2a) due to the varying local strain or twist angles at specific locations. We cannot find the exact correlation between the stripes and the moiré patterns (Methods and Extended Data Fig. 6). Therefore, a conclusion for the origin of the stripes cannot be drawn yet, and further future work is required. Another intriguing property is the distortion of the stripe order mentioned before, which contributes to the decoherence of the long-range order. The possibility of defects acting as possible perturbations of the stripes[37], however, has not been observed in our atomically resolved STM topography (Extended Data Fig. 7).

Next, we discuss the existence and suppression of MBS in the surface stripe region. From the Fu-Kane mechanism[3] we expect MBS to appear in vortex cores when the topological surface states are gapped by the intrinsic superconductivity with a proper bulk chemical potential and pairing strength[11,38]. Tuning these quantities can generate a topological phase transition such that the MBS may disappear (e.g. see ref. [38]). In our case, we have observed the MBS on the bottom layer (with no stripes) of 2M-WS$_2$, indicating that these conditions on the bulk properties can be fully satisfied. We expect that surface modulations should not locally gap out the vortex-bound MBS when vortices are not overlapping laterally on the surface, but our observations indicate their absence. This seeming contradiction is resolved, once we realize that the MBS do survive in the region with the stripe order, but their profile and position are shifted deeper into the bulk away from the top layer by the charge order (Fig. 4b). Moreover, since the superconducting gap away from the surface is only slightly modified compared to the surface itself (see Fig. 3d) we expect that the driving force for the suppression of MBS is the charge order itself on the surface. Such phenomena are in contrast to the observation in the strained LiFeAs, in which the strain-induced bulk charge order gaps the bulk states near the $E_F$ and effectively tunes the chemical potential more favorably for the formation of MBS states in the pinned vortex centre[23]. To understand these phenomenological observations, we build a tight-binding lattice model (see Methods and Extended Data Fig. 8 and Fig. 9) with the non-trivial topological character, and implement both superconductivity and charge order as mean field potentials. The numerical simulations of this model in the presence of a flux tube/vortex show that the MBS can be pushed downwards by surface charge order, when its magnitude is greater than a critical value (see Fig. 4c). This is consistent with our experimental observation, i.e., the absence of MBS on the topmost layer with strong stripe order.

Finally, we discuss the significance of the proposed mechanism of modulating the MBS wavefunction via surface charge order, compared with other accidental mechanisms such as local impurity potentials which was argued as the reason for the absence of MBS in intrinsic topological superconductors[13,21,39]. Using our model, we tested the effects of introducing charge order to more layers near the surface and found that, in contrast to the impurity potential scenario, we can control the location of the MBS within a wide range of depths beneath the surface (see Fig. 4c). This can be understood from the limiting case of inducing charge order on a thick bulk layer near the top surface of the material (Fig. 4d) and analyzing the domain wall problem between the two bulk regimes. On the two sides of the domain wall, the bulk flux tube topology can be characterized by a Zak phase calculation (see Methods), and the resulting detailed phase diagram is shown in Fig. 4e. The MBS is guaranteed to occur at the domain wall if the two sides around the domain wall belong to different topological phases (Fig. 4d). Therefore, the surface charge order observed in our system provides a reliable way to control the MBS's depth by effectively trivializing the surface and creating a domain wall between a topologically trivial phase and nontrivial phase on the flux tube. Our findings not only pave a new path towards manipulating MBSs for quantum information processing and shielding them from spurious external stimuli, but also establish a platform with a rich phase diagram for future studies of complex electronic states in intrinsic topological superconductors.

**Figures and Captions**

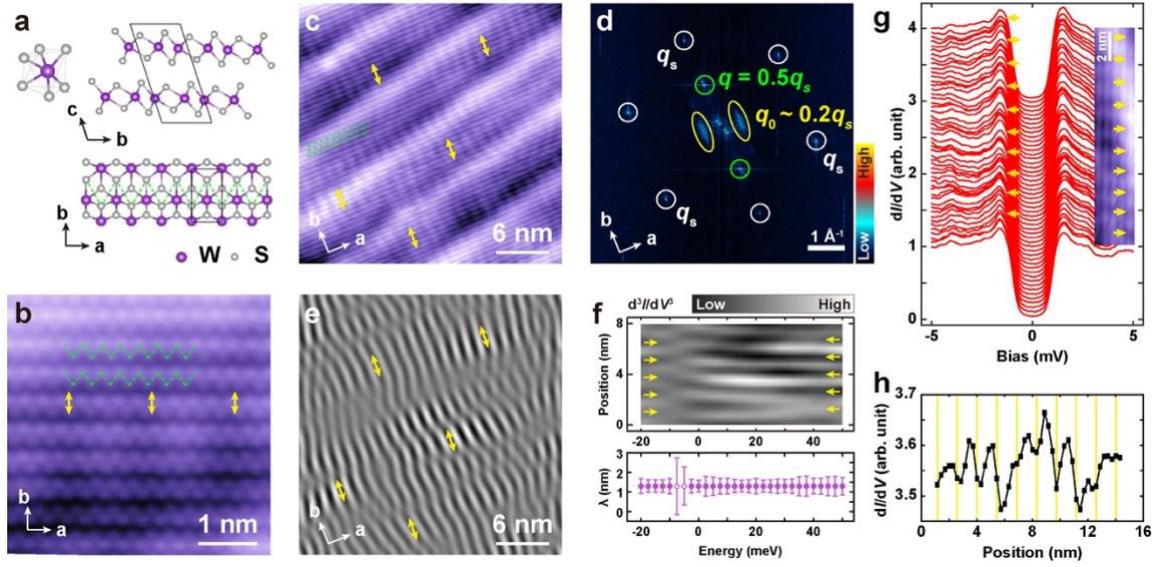

**Fig. 1 | Electronic stripe modulations in 2M-WS$_2$. a,** The atomic structure of 2M-WS$_2$. Top left panel: Schematic of $[WS_6]^{8-}$ structure. Top right panel: Side view of 2M-WS$_2$. The adjacent monolayers stack along the *c*-direction through a translation operation. Bottom panel: Top view of 2M-WS$_2$. Peierls distortion along the *b*-direction leads to a zigzag structure along the *a*-direction denoted by green dashed lines. **b,** Atomically resolved STM topographic image of WS$_2$ (4.4 nm × 4.4 nm; set point, $V_s$ = -5 mV, $I_t$ = 300 pA). The zigzag structure of S atoms is marked by green dashed lines. The double headed yellow arrows denote three valleys of stripes in this image. **c,** Stripe modulations over a large area of WS$_2$ (30 nm × 30 nm; set point, $V_s$ = -5 mV, $I_t$ = 100 pA). The large diagonal features running from bottom left to top right correspond to moiré patterns. The green dashed lines denote three adjacent zigzag chains. The double headed yellow arrows in **c** and **e** denote some valleys of stripes. **d,** Fast Fourier transform result of **c**. The two parallel, line-features (near $q_0$) correspond to the stripe modulations in real space. The wavevectors $q$ and $q_s$ correspond to the zigzag chains of S atoms and the S-lattice, respectively. The length of $q_0$ (~ 0.2$q_s$) is estimated by the wavevector normal to the line-features. **e,** Inverse fast Fourier transform of $q_0$. The stripe modulations are extracted and clearly displayed. **f,** Top panel: A line-cut across several stripes of the second-order derivative of d$I$/d$V$ as a function of energy. The modulations show a non-dispersive behavior in energy. The yellow arrows denote the valleys of in d$^3I$/d$V^3$ this panel. Bottom panel: The period of the stripes as a function of energy, which is determined from the peak position in the FFT result of the d$^3I$/d$V^3$ line-cut at each energy. The error-bar is extracted from the peak width in the FFT. The data labelled in hollow purple circles are in the transition energy ranges, in which the signals are relatively weak. **g** and **h,** Spatially periodic modulations of the superconducting coherence peaks. **g,** A series of low-energy d$I$/d$V$ spectra (set point, $V_s$ = -5 mV, $I_t$ = 400 pA) taken in right inset. Right inset: STM topography of WS$_2$ (2.8 nm × 14.6 nm; set point, $V_s$ = 10 mV, $I_t$ = 100 pA). The yellow arrows denote the valleys of the stripe patterns (inset) and the corresponding locations where the spectra are acquired. **h,** Position dependence of averaged height of the symmetric coherence peaks in **g**, which shows clear periodic spatial modulation. The height of

individual coherence peaks is determined by the average of five nearest neighbours near the maximum. The vertical yellow lines correspond to the positions of yellow arrows in the inset of **g**.

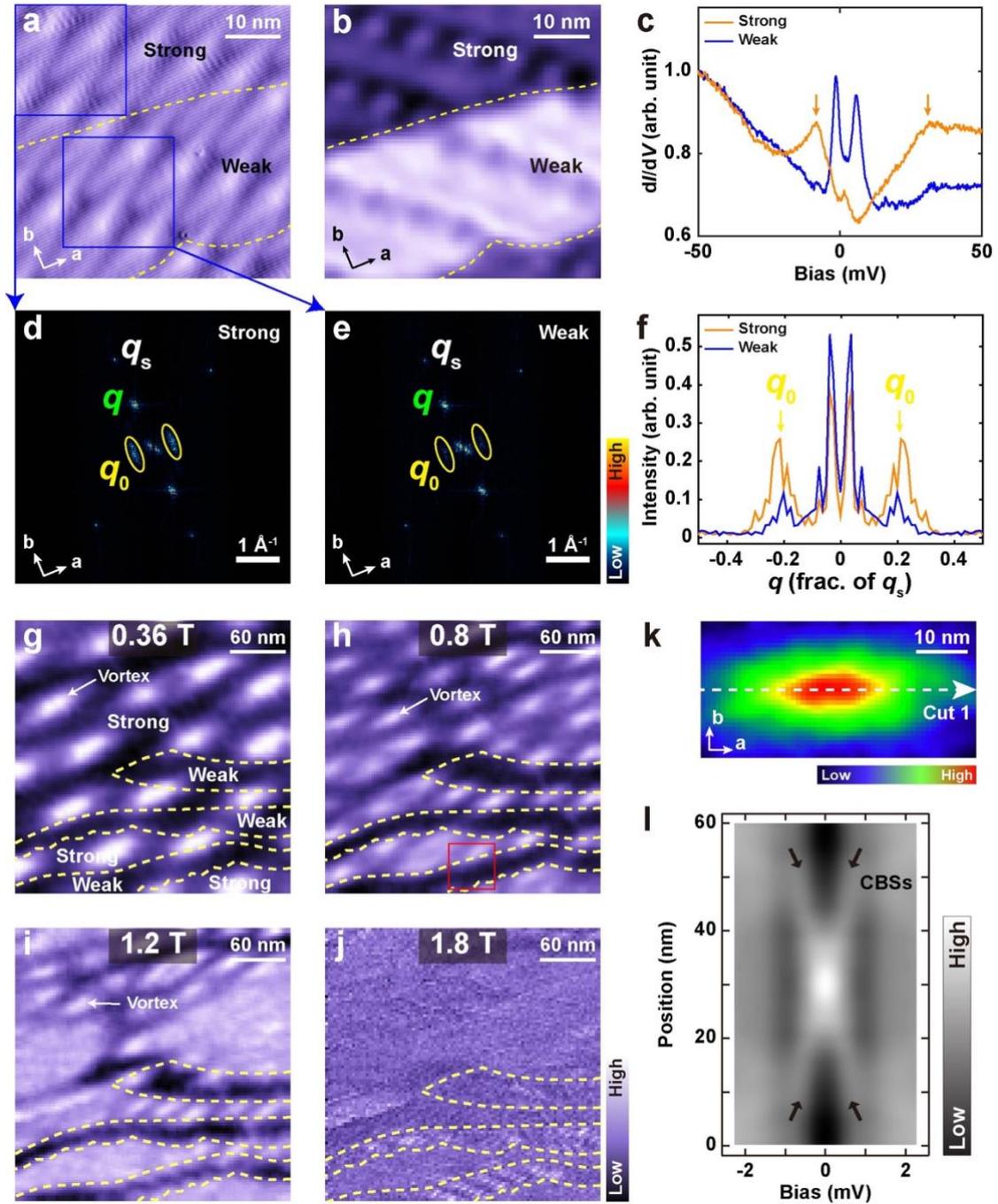

**Fig. 2 | Spatial distribution of magnetic vortex lattice and absence of Majorana bound state in vortex core of WS$_2$ with stripes. a**, Derivative along the $x$-axis of a topographic image of WS$_2$ (50 nm × 50 nm; set point, $V_s$ = -10 mV, $I_t$ = 100 pA), in which two types of stripe modulations are displayed. The yellow dashed lines denote the boundaries between regions with strong and weak stripes. The monoclinic features in large periodicity in this panel are moiré patterns and will be discussed in Discussion section. **b,** d$I$/d$V$ mapping at 5 mV of the same area in **a** at 12 T (set point, $V_s$ = -20 mV, $I_t$ = 200 pA). The region with higher intensity of DOS corresponds to weak stripe region. **c,** Typical d$I$/d$V$ spectra of the two regions under magnetic field of 3 T (set point, $V_s$ = -50 mV, $I_t$ = 500 pA). The density of states (DOS) near the Fermi level ($E_F$) in weak stripe region is much higher than those in strong stripe region. A gap-

like feature (denoted by the yellow arrows) appears in the spectrum of the strong stripe region. **d** and **e,** Fast Fourier transforms of the strong and weak stripe regions in **a** (marked by blue boxes), respectively. The intensities of the two FFT images are normalized by their zigzag signal $q$. **f,** Line profiles cutting across $q_0$ in **d** and **e**. The $x$-axis is in unit of $q_s$. **g-j,** Zero-bias conductance maps of $WS_2$ (300 nm × 300 nm; set point, $V_s = -5$ mV, $I_t = 200$ pA) taken at 0.36 T (**g**), 0.8 T (**h**), 1.2 T (**i**) and 1.8 T, above $H_{c2}$ (**j**), respectively. Magnetic vortices appear in only the strong stripe regions and their density increases with higher magnetic fields. The red box in **h** shows where the topographic image of **a** is taken. **k,** A zero-bias conductance map (60 nm × 30 nm; set point, $V_s = -5$ mV, $I_t = 200$ pA) of a single vortex at 0.36 T. The vortex is anisotropic and elongated along the *a*-direction. **l,** False-colour image of a series of d$I$/d$V$ spectra (set point, $V_s = -4$ mV, $I_t = 400$ pA) measured along cut 1 in **k**. Along the *a*-direction, the single zero-bias conductance peak splits into two symmetric branches as the tip moves away from the vortex core, corresponding to Caroli-de Gennes-Matricon bound states (CBSs).

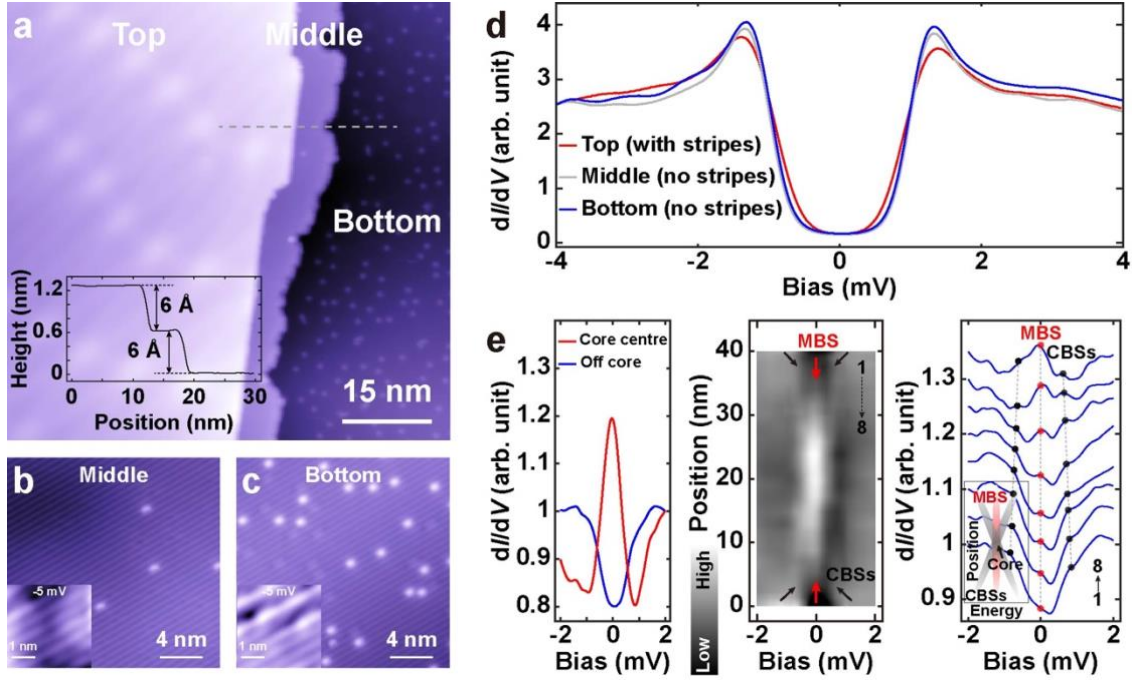

**Fig. 3 | Recovered Majorana bound states in the underlying WS$_2$ layer with no stripes. a**, Large-scale topographic image of WS$_2$ (75 nm × 75 nm; set point, $V_s$ = -2 V, $I_t$ = 20 pA) that includes three layers of WS$_2$. The top layer is the region with the stripe patterns studied in Fig. 1-3. Inset: STM topographic scan profile taken across the two steps, denoted by the gray dashed line in **a**, and the step heights are both 6 Å, corresponding to the *c*-axis lattice constant of 2M-WS$_2$. **b-c,** Topographic images of the middle (20 nm × 20 nm; set point, $V_s$ = -1 V, $I_t$ = 50 pA) and bottom (20 nm × 20 nm; set point, $V_s$ = 1 V, $I_t$ = 20 pA) layers, respectively. Insets: zigzag chains of S atoms (4 nm × 4 nm; set point, $V_s$ = -5 mV, $I_t$ = 300 pA). The stripes are absent in the both layers. **d,** Averaged d$I$/d$V$ spectra of the three layers (set point, $V_s$ = -4 mV, $I_t$ = 200 pA). Compared with the top layer, the superconducting gaps are enhanced in the middle and bottom layers. The unaveraged spectra are shown in Extended Data Fig. 3. **e,** A series of d$I$/d$V$ spectra taken along the elongated direction (set point, $V_s$ = -4 mV, $I_t$ = 200 pA) of a magnetic vortex on the bottom layer at 0.36 T. Left panel: d$I$/d$V$ spectra measured on vortex core (red curve) and away from core (blue curve). False-colour image of the spectra (middle panel) demonstrates three branches of bound states denoted by red and black arrows in the vicinity of the vortex. The splitting branches are attributed to CBSs, and the robust non-split branch at $E_F$ is a Majorana bound state. The red and black dots on the spectra (right panel) indicate the three-peak feature and the spatial evolutions of Majorana bound state and CBSs off the vortex core. Inset: schematic of the evolution of the coexisting Majorana bound state and CBSs.

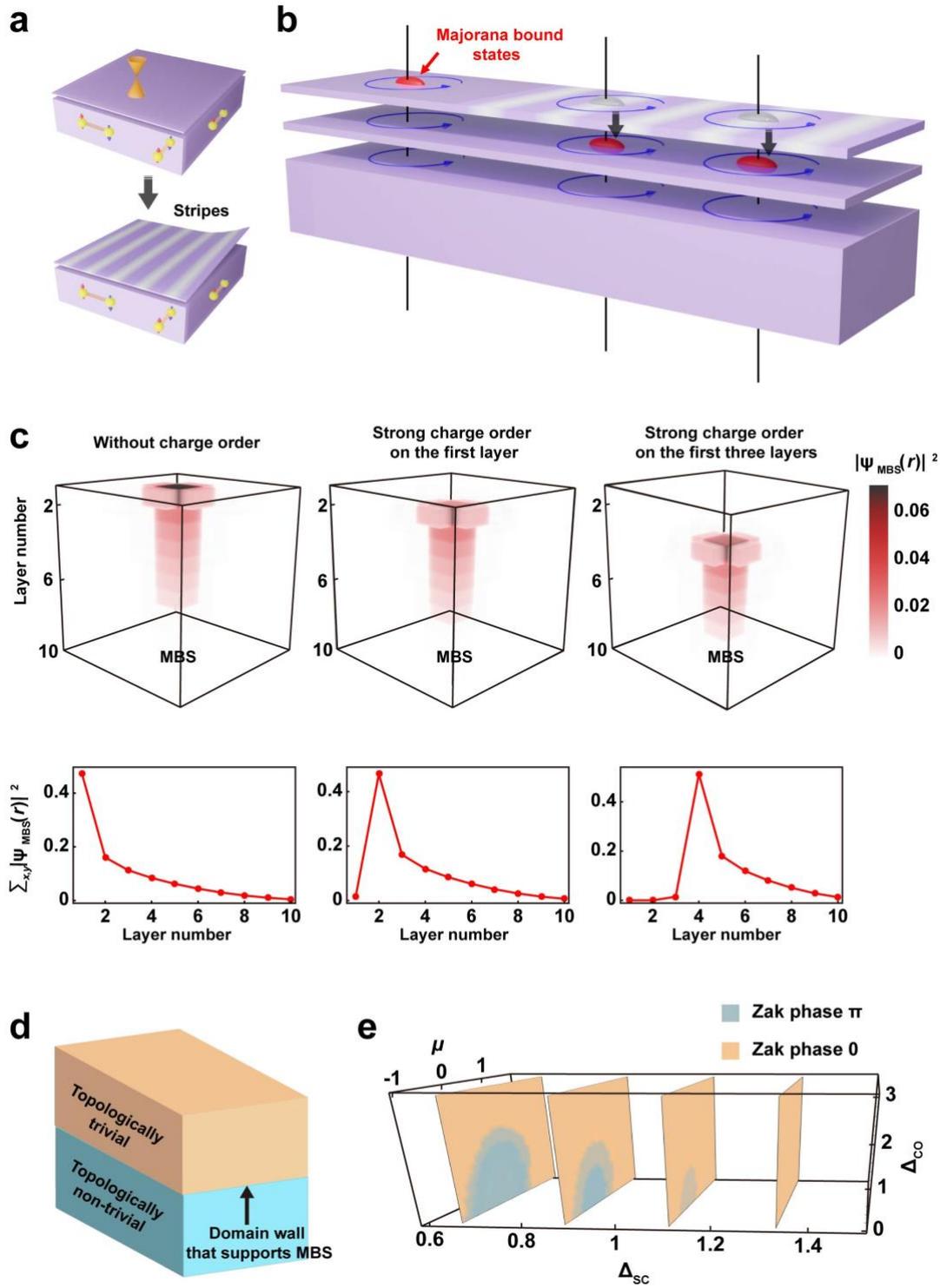

**Fig. 4 | Suppression of Majorana bound states by the surface stripe order. a,** Development of the stripes during the sample cleavage. **b,** Suppression of MBS in the surface stripe region. MBS are pushed downward to the layer beneath the topmost layer with stripe order. **c,** Tight-binding lattice calculation results for the norm-squared of the MBS wavefunctions ($|\psi_{MBS}(\mathbf{r})|^2$) on the top surface without and with strong surface charge order. This quantity corresponds to the probability of finding an electron/hole at different positions. The top three panels show the

real space distribution of $|\psi_{MBS}(\mathbf{r})|^2$ for the MBS, and the bottom three panels show the norm-squared of wavefunctions along the normal direction of the layers, corresponding to $\sum_{x,y}|\psi_{MBS}(\mathbf{r})|^2$. Notice that we also include cases of adding charge order to multiple layers near the surface. **d,** A domain wall created from adding an extensive number of layers with strong charge order on the top surface. In contrast to other local effects such as a random impurity potential, adding charge order can push the MBS to any depth due to the formation of the domain wall that supports MBS. **e,** Bulk phase diagram in parameter space spanned as a function of the chemical potential $\mu$ relative to the Fermi-level, the charge order strength $\Delta_{CO}$, and the superconductivity pairing strength $\Delta_{SC}$ from tight-binding calculations. The details of the tight-binding model and the convention of the energy unit can be found in the Methods part. The blue region has Zak phase $\pi$ (topologically nontrivial) and the yellow region has Zak phase 0 (topologically trivial). The two sides of the domain wall highlighted in **d** correspond to the two phases in **e**.

## Methods

### Sample preparation

2M WS$_2$ crystals were synthesized by a topochemical method. To synthesize the precursor K$_{0.7}$WS$_2$, the reactants W, S, and K$_2$S$_2$ powders were mixed in a stoichiometric ratio and pressed into a pellet in an argon glove box, which was sealed in an evacuated silica tube at 10$^{-5}$ Torr. The tube was heated up to 850 °C for 5°C/min, maintained at this temperature for 6000 minutes, and cooled to 600 °C at a rate of 0.1 °C/min in the muffle tube. The as-synthesized K$_{0.7}$WS$_2$ (0.1 g) crystals were dispersed in de-ionized water and stirred in the acidic K$_2$Cr$_2$O$_7$ (0.01 mol/L) aqueous solution for 1 hour at room temperature. Finally, the 2M-WS$_2$ crystals were obtained after washing them in distilled water several times and drying them in the vacuum oven.

### Crystal structure determination

The crystal structure of 2M WS$_2$ crystals was determined through the refinement of the single-crystal X-ray diffraction in the program SHELXS. The chosen suitable-size 2M WS$_2$ single-crystal was set in a Bruker D8 Quest diffractometer equipped with Mo Kα radiation to collect the diffraction data at room temperature. The absorption corrections were performed using the multiscan method. The detailed crystallographic data and refinement details of 2M WS$_2$ were shown in our previous paper [ref. 24].

### STM measurements

Our experiments were performed on an ultra-high vacuum (UHV) commercial STM system (Unisoku) which reaches a base temperature of 400 mK by using a single-shot $^3$He cryostat. The base pressure of the system is 2.0×10$^{-10}$ Torr. Crystalline WS$_2$ samples were cleaved *in-situ* at 78 K then transferred into STM. All the measurements are done at 400 mK. A polycrystalline PtIr STM tip was used and calibrated using Ag island before STM experiments. STS data were taken by standard lock-in method. The feedback loop is disrupted during data acquisition and the frequency of oscillation signal is 973.0 Hz.

### Estimation of $H_{C2}$

We estimate the $H_{C2}$ by analyzing the tunneling spectra acquired from areas away from vortices under different magnetic fields. We extracted zero-bias conductance (ZBC) in the spectra and summarized in Extended Data Fig. 1. The linear fitting results suggest that the upper critical field is 1.6 T (1.9 T) in strong (weak) stripe region. The higher upper critical field in the weak stripe region may originate from the size effect or indicates the competing between stripes and superconductivity.

### Distribution of stripes under magnetic field

To address the magnetic field dependence of the distorted stripes in 2M-WS$_2$, differential topographic images of the same area are investigated (Extended Data Fig. 2) under 0 T and 12 T, respectively. The distributions of the stripes in the two images are identical, exhibiting that the stripes are not sensitive to the external magnetic field.

### First-principles calculations

We performed *ab initio* DFT calculations to obtain the electronic band structures and phonon dispersions for monolayer, bilayer and bulk 2M-WS$_2$. The electronic band structures are shown in Extended Data Fig. 4. Our results are consistent with the previous studies[15,33]. The bulk 2M-WS$_2$ is topologically non-trivial with inverted band structures around the Γ point. Extended Data Figure 5 shows the calculated phonon spectrums under three doping conditions, neutral, electron doped and hole doped for monolayer, bilayer, and bulk 2M-WS$_2$, respectively. The red dashed lines mark the wave vector $q_0$ for the stripe charge order (approximately 0.48 Å$^{-1}$ along the direction of zig-zag chains) observed in STM. As shown in Extended Data Fig. 5, we do not observe the softening of any phonon modes around $q_0$. Therefore, the electron-phonon interaction cannot be the origin of the stripe order.

First-principles calculations were performed with the Vienna *ab initio* simulation package (VASP). The projector augmented wave (PAW) was used to describe the core level of atomic potential[40,41]. The generalized gradient approximation (GGA) developed by Perdew-Burke-Ernzerhof[42] was used for the exchange-correlation functional. A plane-wave cutoff of 450 eV was used for the wavefunctions. For the self-consistent electronic structure calculations, we set the energy convergence criterion as 10$^{-6}$ eV. The bulk lattice structures were fully relaxed until the force is smaller than 0.001 eV/Å. For the monolayer and bilayer structures, a vacuum layer of 20 Å were adopted along the $z$ axis in the supercell to avoid the coupling between neighboring supercells. The phonon band structures for monolayer, bilayer and bulk 2M-WS$_2$ were calculated by using the PHONOPY package[43]. The spin−orbit coupling effect was included in these calculations for electronic structures, but was absent for the phonon calculations. For the calculations, the Brillouin zones (BZs) were sampled by 16 × 8 × 1 k-grid for monolayer and bilayer and 8 × 8 × 10 for bulk. For phonon calculations, we adopted 4 × 2 × 1 supercells for monolayer and bilayer 2M-WS$_2$ thin film and 2 × 2 × 4 for bulk 2M-WS$_2$.

**The cross-correlation analysis of the stripes and moiré patterns**
A recent study on ZrTe3 has shown that local impurities are able to apply strong pinning potential and thus introduce distortion and phase modulation to CDW[37]. Since the moiré patterns and distorted stripe charge order coexist in WS$_2$, one natural question would be whether the moiré patterns play the similar role. The following cross-correlation analysis excludes such scenario.

We first extract the stripe and moiré distribution from the topographic image (Extended Data Fig. 6). Extended Data Fig. 6b presents the stripe distribution obtained by taking the inverse fast Fourier transform (IFFT) of the stripe wave vectors in the FFT of Extended Data Fig. 6a. Extended Data Fig. 6c is the AA sites of moiré patterns in Extended Data Fig. 6a, which is obtained by collecting the maximum DOS value in Extended Data Fig. 6a. The two images are normalized by the following formula:

$$F(x,y) = \frac{f(x,y) - \mu}{\sqrt{n}\sigma}$$

Here, $F(x,y)$ is the normalized distribution, $f(x,y)$ is the raw distribution, $n$ is the pixel number of $f$, $\mu$ and $\sigma$ are the mean value and standard deviation of $f$, respectively.

Then the normalized cross-correlation is calculated using the formula below

$$[F * G](x, y) = \sum_{x_0, y_0} F(x_0, y_0) G(x_0 + x, y_0 + y)$$

Here, $[F * G](x, y)$ is the cross-correlation between $F$ and $G$ with displacement of $(x, y)$.

Extended Data Fig. 6d shows the cross-correlation result between Extended Data Fig. 6b and c. Here we zoom-in the displacement to (±10 nm, ±10 nm) region, in the order of single moiré unit cell. The direct onsite influence from the AA site on stripes is easily excluded since the cross-correlation at (0 nm, 0 nm) displacement is negligible. The non-local influence, from long-range interaction or AB site, is also excluded for two reasons: (1) the maximum value in Extended Data Fig. 6d is very small, only about 0.02. This indicates the influence from finite displacement is also negligible. For clarity, a line profile taken along the black arrow is shown in Extended Data Fig. 6e to present the absolute cross-correlation value. (2) The same cross-correlation analysis to another topographic image (Extended Data Fig. 6f) shows a totally different distribution, which indicates the local maxima in the cross-correlation are just from random fluctuations rather than the exact interactions between moiré pattern and stripes.

**Tight-binding calculations**

To capture the nontrivial topology and the resulting MBS on the surface of 2M-WS$_2$, we can study a simple Fu-Kane type tight-binding model[3], which is a 3D topological insulator (TI) in a cubic lattice with s-wave superconducting pairing. We conduct calculations for such a tight-binding model to show that (i) the surface charge order can indeed suppress the MBS from the top surface, and (ii) the suppression of the MBS in position space can be understood from the bulk flux tube topology with the bulk superconductivity and nontrivial topological surface states.

The Bogoliubov-de-Gennes (BdG) Hamiltonian of our tight-binding model is

$$H_{\text{BdG}} = \frac{1}{2} \sum_{k} \Psi_{\mathbf{k}}^{\dagger} [t \sin k_x \xi_z \tau_z \sigma_x + t \sin k_y \xi_z \tau_z \sigma_y + t \sin k_z \xi_z \tau_y \sigma_0$$

$$+ (m + \sum_{i=x,y,z} t \cos k_i) \xi_z \tau_x \sigma_0 + (V(\mathbf{k}) - \mu) \xi_z \tau_0 \sigma_0 + \Delta_{SC} (\cos \phi \, \xi_x \tau_0 \sigma_0 + \sin \phi \, \xi_y \tau_0 \sigma_0)] \Psi_{\mathbf{k}}$$

where $k_{x,y,z}$ are the momenta along $x, y, z$ directions, $\xi_{x,y,z}$ are the Pauli matrices for the particle and hole degree of freedom, $\tau_{x,y,z}$ are Pauli matrices for the atomic orbital, and $\sigma_{x,y,z}$ are Pauli matrices for the spin. $\xi_0, \tau_0$, and $\sigma_0$ are 2 × 2 identity matrices. The electrons are considered to be living on discrete atomic sites and the $t$ is the hopping strength among them. For simplicity, we set $t = 1$ in the following. $\mu$ is the chemical potential. $\Delta_{SC} = \Delta_0 e^{i\phi}$ is the s-wave superconductivity pairing parameter. $m$ is the mass term that controls the topology of the insulating ground state in the absence of the superconductivity and charge order potential: i.e., if we take $-3 < m < -1$, then without superconductivity pairing and charge order potential, the model is a 3D TI that has one surface Dirac cone on each surface. $\Psi_{\mathbf{k}}$ is the

Nambu basis that takes the form

$$\Psi_{\mathbf{k}} = \begin{pmatrix} c_{\mathbf{k},1\uparrow} & c_{\mathbf{k},1\downarrow} & c_{\mathbf{k},2\uparrow} & c_{\mathbf{k},2\downarrow} & c^{\dagger}_{-\mathbf{k},1\downarrow} & -c^{\dagger}_{-\mathbf{k},1\uparrow} & c^{\dagger}_{-\mathbf{k},2\downarrow} & -c^{\dagger}_{-\mathbf{k},2\uparrow} \end{pmatrix}^{T}$$

Where 1 and 2 (up and down arrows) label orbital (spin) degrees of freedom. V($\mathbf{k}$) is a momentum-dependent potential function. With zero V($\mathbf{k}$), the model is insulating in the bulk, and with nonzero V($\mathbf{k}$), the model generally has no direct gap in the energy spectrum, which is more closed to the experimental situations. However, since the bulk Fermi surface does not qualitatively affect the physics (topology) related to the existence and distributions of MBS, we set V($\mathbf{k}$) = 0 for results shown in the main text, and relegate the calculations with nonzero V($\mathbf{k}$) in Extended Data Fig. 8.

To directly observe charge-order induced suppression of MBS from the top surface, we Fourier transform the above Bloch Hamiltonian into real space and calculate the eigenstate wavefunctions for a $10 \times 10 \times 10$ lattice with open boundary conditions along all three directions. We introduce a vortex in the *x-y* plane (i.e., a flux tube along the *z*-direction) by assigning different phases of pairing parameters, $\phi$, in different patches of the lattice, as illustrated in Extended Data Fig. 9.

We label the ten discrete atomic layers in our tight-binding model along *z* direction by $z = 1, 2, \ldots, 10$, where $z = 1$ corresponds to the top layer and $z = 10$ corresponds to the bottom layer. We add the surface charge order potential on the layer at $z = 1$ as $\Delta_{CO} \cos(Qx) \delta_{z,1} \xi_z \tau_0 \sigma_0$, where $\delta_{z,1}$ is the Kronecker Delta function. When $\Delta_{CO}$ is large enough, we can observe the suppression of MBS from the top surface to the deeper region as discussed in the main text. Without loss of generality, in all our calculations, we use $m = -2t$, $\Delta_0 = t$, and $Q = 2\pi/3$. The choice of $Q$ will not influence the suppression of MBS from the top surface induced by the surface charge order (see Extended Data Fig. 10). For the plots shown in Fig. 4c, we fix $\mu = 0$, and choose $\Delta_{CO} = 10t$ ($\Delta_{CO} = 0$) for the second and third (first) panel. Since we are doing calculations in a finite system, there are no exact zero modes - MBS localized on opposite surfaces due to the hybridization between each other. We use a symmetric linear combination for two eigenstates, which are closest to zero energy in our calculations. Then we plot the symmetric results as the probability of the MBS localized on the top surface (i.e., the surface at *z* = 1 as shown in Fig. 4c).

Interestingly, if we switch on charge order deeper in the bulk, the MBS is suppressed further into the bulk and appears at the interface between regions with and without charge order (see Fig. 4c). Consequently, a global downward displacement of the entire MBS wavefunction can be observed with the appearance of surface charge order on the top layers (see Fig. 4c). This is because, with the Nambu basis and considering the superconducting pairing, the charge order can lead to a bulk topological phase transition in the 1D flux tube (see Fig. 4e), and the interface between regions with and without charge order is a domain wall between distinct topological phases. At the interface (see Fig. 4d), the MBS is trapped. Technically, the bulk topological property for each region on opposite sides of the domain wall (marked as blue and yellow in Fig. 4d) could be captured by the Zak phase along *z*-direction (i.e., the direction of the flux tube) computed in the translationally invariant bulk, respectively. Conventionally, a system with Zak phase $\pi(0)$ is identified as topologically nontrivial (trivial), which hosts (no) MBS on its boundary. In order to calculate the Zak phase, instead of having all three directions open, we need to keep the $z$-direction periodic so that the momentum along the $z$-direction, $k_z$, is a good quantum number. Furthermore, our model has a mirror-$z$ symmetry ($M_z = \xi_0 \tau_x \sigma_z$) and this simplifies the calculation of Zak phase along $z$ direction as the product of eigenvalues of the mirror-$z$ operator for occupied eigenstates at $k_z = 0$ and $k_z = \pi$. Given a chemical potential $\mu$, a charge order parameter $\Delta_{CO}$, and a superconductivity pairing order parameter $\Delta_{SC}$, we can calculate the corresponding Zak phase. By scanning these parameters, we plot the phase diagram in Fig. 4e.

**Data availability**
The data and analysis supporting the findings of this paper are available from the corresponding authors upon reasonable request.

**Acknowledgements** We thank Y.Y. Wang, Z.Y. Weng and W.H. Duan for helpful discussions. The experimental work was supported by the National Science Foundation (Grants No. 11427903, No. 51788104, No. 12234011) and Ministry of Science and Technology of China (Grants No. 2022YFA1403100). W. Li was also supported by Tsinghua University Initiative Scientific Research Program, Beijing Young Talents Plan and the National Thousand-Young-Talents Program. P. Tang was supported by the Open Research Fund Program of the State Key Laboratory of Low-Dimensional Quantum Physics. For the theory work P. Zhu and T.L. Hughes were supported by ARO MURI W911NF2020166 and X.Q. Sun was supported by the Gordon and Betty Moore Foundation's EPiQS initiative through Grant GBMF8691.

**Author contributions** W.L. and Q-K.X. designed and coordinated the experiments; X.F. and Y.Y. did the STM experiments; Y.F. and F.H. grew the samples; X.S., P.Z. and T.H. performed the tight-binding simulation and analyzed the theoretical results. P.T. and Y.J. carried out the DFT calculations. W.L., X.F. and Y.Y. analyzed the experimental data. W.L., X.F., P.T. and X.S. wrote the manuscript with comments from all authors.

**Competing financial interests** The authors declare no competing financial interests.